\documentclass{emulateapj}

\newcommand{\Kepler}{{\it Kepler}}

\newcommand{\rearth}{R$_\oplus$}

\newcommand{\kms}{\ensuremath{\rm km\,s^{-1}}}
\newcommand{\ms}{\ensuremath{\rm m\,s^{-1}}}

\newcommand{\ntargets}{59,174}
\newcommand{\nplanets}{234}
\newcommand{\nsys}{208}
\newcommand{\nst}{68}
\newcommand{\nmulttres}{26}
\newcommand{\nspec}{101}

\usepackage{hyperref}
\usepackage{breakurl}
\usepackage{amsmath}

\slugcomment{}

\shorttitle{K2 Planet Candidates}
\shortauthors{Vanderburg et al.}

\begin{document}


\title{Planetary Candidates from the First Year of the K2 Mission}


\author{Andrew Vanderburg\altaffilmark{$\dagger$ 1}, David W. Latham\altaffilmark{1}, Lars A. Buchhave\altaffilmark{2}, Allyson Bieryla\altaffilmark{1}, Perry Berlind\altaffilmark{1}, Michael L. Calkins\altaffilmark{1}, Gilbert A. Esquerdo\altaffilmark{1}, Sophie Welsh\altaffilmark{1}, \& John Asher Johnson\altaffilmark{1}}


\altaffiltext{$\dagger$}{\url{avanderburg@cfa.harvard.edu}}
\altaffiltext{1}{Harvard--Smithsonian Center for Astrophysics, 60 Garden St., Cambridge, MA 02138}
\altaffiltext{2}{Centre for Star and Planet Formation, Natural History Museum of Denmark \& Niels Bohr Institute, University of Copenhagen, \O ster Voldgade 5-7, DK-1350 Copenhagen K.}


\begin{abstract}
The \Kepler\ Space Telescope is currently searching for planets transiting stars along the ecliptic plane as part of its extended K2 mission. We processed the publicly released data from the first year of K2 observations (Campaigns 0, 1, 2, and 3) and searched for periodic eclipse signals consistent with planetary transits. Out of \ntargets\ targets we searched, we detect \nplanets\ planetary candidates around \nsys\ stars. These candidates range in size from gas giants to smaller than the Earth, and range in orbital periods from hours to over a month. We conducted initial reconnaissance spectroscopy of \nst\ of the brighter candidate host stars, and present high resolution optical spectra for these stars. We make all of our data products, including light curves, spectra, and vetting diagnostics available to users online.  

\end{abstract}


\section{Introduction}

NASA's \Kepler\ Space Telescope is a Discovery class spacecraft designed to search for transiting exoplanets with simultaneous precise photometric measurements of over 100,000 stars in one 115 square degree field of view. \Kepler\ has been highly successful, finding thousands of planetary candidates \citep{koi2, koi6}, the majority of which are likely genuine transiting exoplanets \citep{morton,morton12, fressin, lissauer, rowe, pastis}. \Kepler's success came from its extreme photometric precision, enabled in part by four gyroscope-like reaction wheels which stabilized the spacecraft's pointing. When the second of the four reaction wheels failed in May 2013, \Kepler\ was left unable to point precisely at its original field. The spacecraft was, however, able to balance itself against solar radiation pressure and continue taking data with only two reaction wheels in its new K2 mission, albeit with significantly worsened pointing precision. In K2 operations, \Kepler\ observes fields pointed away from the sun, along the ecliptic plane, for only about 80 days at a time before moving onto an entirely new field \citep{howell}. 

Initial characterization of 9 days of K2 data showed that the reduced pointing precision introduced significant levels of systematic noise into the light curves, and that without special processing, K2 data could attain photometric precision within about a factor of four of typical \Kepler\ data on six hour timescales. \citet[hereafter VJ14]{vj14} subsequently showed that the systematics introduced by K2's reduced pointing precision were tightly correlated with the motion of the spacecraft. By decorrelating the spacecraft's motion with the measured photometry, VJ14 were able to improve the quality of K2 photometry to be close (within 35\% for bright stars) to the precision achieved by \Kepler. Soon thereafter, K2 began successfully detecting new exoplanets \citep{hip116454,crossfield}. Since then, a variety of new K2 data analysis techniques and planet detections have been presented \citep{aigrain, k2varcat, armstrong2, lacourse, foremanmackey, montet, roberto, huangk2c1, huang, angus, lund}. Despite the proliferation of K2 pipelines and planet searches, a uniform search for planet candidates across multiple campaigns has not yet been presented.

In this paper, we undertake a search for transiting planet candidates from Campaigns 0, 1, 2, and 3 of K2 data. In Section \ref{k2data}, we describe the characteristics of the K2 data and our photometric analysis technique, including extracting light curves from pixel level data and correcting for systematics introduced by \Kepler's unstable pointing. In Section \ref{vetting}, we describe our transiting planet search and the procedures we use to vet the results of our search to sort out spurious detections and astrophysical false positives. In Section \ref{transitfits}, we describe the process with which we model the transits to extract planet properties, and in Section \ref{stellarparams}, we estimate stellar parameters of candidate planet hosts and present the results of initial reconnaissance spectroscopy of some of the brighter candidate host stars. 

\section{K2 Data}\label{k2data}

\subsection{K2 Data Characteristics}\label{data}

In this paper, we search for transiting planets in data collected during Campaigns 0, 1, 2, and 3 of the K2 mission. During this time period after the initial engineering test proof of concept, the \Kepler\ team was still learning about the performance of the spacecraft, and as time went on and the team's confidence in the spacecraft grew, the characteristics of the data changed. We focus our efforts on data collected of individual ``Guest Observer'' targets, and ignore targets observed as part of ``super-stamps'' and targets observed serendipitously as part of the target masks for other objects. 

\subsubsection{Campaign 0}
During Campaign 0, K2 observed a field centered at RA = 6:33:11.14, Dec = +21:35:16.40, for a period of 80 days between March and May of 2014. This initial campaign was designed to be a full length shakedown of the K2 operating mode, so relatively large pixel masks were downloaded surrounding each target. During Campaign 0, Jupiter passed nearby the field (drifting across a dead CCD module). Reflections from Jupiter caused the spacecraft to lose fine-pointing control during the first half of the campaign, so in this work we ignore the first half of the campaign and focus on the last 33 days of data, from BJD = 2456772 to BJD = 2456805. Jupiter's reflections onto the focal plane also introduced systematics into light curves of some stars. In particular, a spike in background scattered light happened when Jupiter entered and exited the focal plane. Improper background subtraction sometimes led to spikes and dips in the light curves at this time. In other cases, Jupiter's reflection passes near or into target's pixel masks, leading to high amplitude systematics. 

Campaign 0 was centered near the plane of the Milky Way galaxy, and as such, the density of stars in this field is quite high, and contamination with distant giant stars is higher than in other campaigns. Also, the typical stars observed tended to be hotter than those observed in the original \Kepler\ mission and other K2 campaigns. These factors significantly increase the false positive rate for candidates in this field. 

\subsubsection{Campaign 1}

During Campaign 1, K2 observed a field centered at RA = 11:35:45.51, Dec = +01:25:02.28 for 83 days between June and August of 2014. We ignore data taken during the first two days of observations, when K2 was pointed about 2 pixels away from its final position. There is a 3 day gap in the data in the middle of this campaign as \Kepler\ pointed away from its field to download data to Earth.  Field 1 is situated at the Northern galactic pole, so the density of stars in this field is quite low, leading to high data quality and a low false positive rate. 

\subsubsection{Campaign 2}

Field 2 of the K2 mission is centered at RA = 16:24:30.34, Dec = -22:26:50.28, and was observed for 79 days between August and November 2014. There was no midcampaign data download, so observations were nearly continuous over the 79 day campaign. Halfway through the campaign, however, K2 switched the direction of its roll, and we ignore about 14 hours of data while the spacecraft drifted to its new equilibrium. During Campaign 3, Mars passed into the K2 field of view, but did not cause as many problems as Jupiter did in Campaign 0. Field 2 is pointed towards the Rho Ophiucus cloud complex, and as such there is significant reddening in this field. This complicated target selection, and led to high rates of giant star contamination in the nominal dwarf star targets. 

\subsubsection{Campaign 3}

Field 3 of the K2 mission is centered at RA = 22:26:39.68, Dec = -11:05:47.99, and was observed for 69 days between November 2014 and February 2015. We ignore the first day of data when K2 was settling into its roll equilibrium. Once again, there was no midcampaign data download, but the drift period when K2 switched the direction of its roll was not as significant as in Campaign 2, so we don't exclude any data midcampaign. A significant difference between Campaign 3 and all previous campaigns is that the timescale for spacecraft attitude control adjustments was decreased from 50 seconds in Campaigns 0-2 to 20 seconds in Campaign 3 (compared to 10 seconds for the original \Kepler\ mission). This decreased the motion of the spacecraft during exposures, which led to an improvement in photometric precision (see Section \ref{c3precision}).

\subsection{Light Curve Processing}\label{lightcurves}

\begin{figure*}[t!]
\epsscale{1}
  \begin{center}
      \leavevmode
\plotone{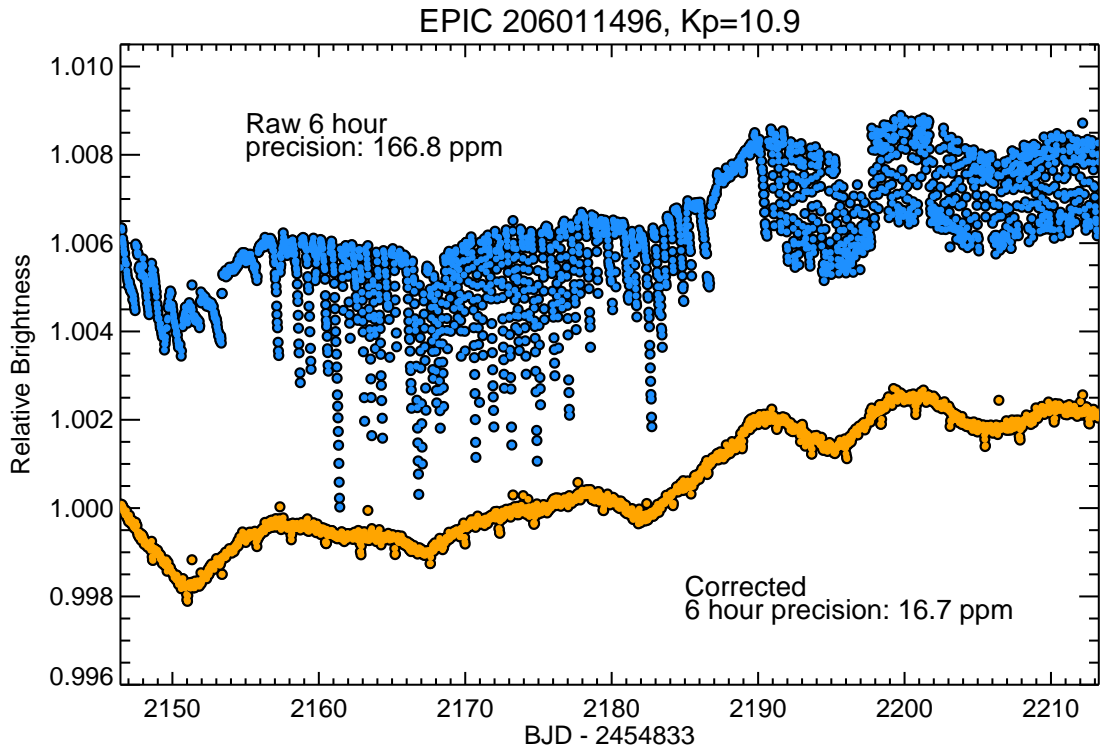}
\caption{Raw (top, blue) and corrected (bottom, orange) light curves of the 11th magnitude star EPIC 206011496 observed by K2 during Campaign 3. The systematics correction greatly improves the data quality and makes it feasible to search for shallow transits that are much smaller than the level of the roll systematics. Periodic transits are visible in the corrected light curve of this particular star, evidence of a 2.37 day period candidate super-Earth. The corrected light curve also shows evidence for a $\simeq$ 14 day rotation period. } \label{sff}
\end{center}
\end{figure*}

After the end of campaigns, the K2 data were downloaded to Earth, and the K2 pixel level data were processed with the calibration module (CAL) of the \Kepler\ pipeline \citep{jenkins}. About 3-4 months after the end of each campaign, the K2 pixel level data were released to the public, at which time we downloaded the target pixel files for all targets from the Mikulski Archive for Space Telescopes (MAST). We processed the data in as uniform a manner as possible given the differences between the various campaign datasets using a descendant of the pipeline described in VJ14. 

\subsubsection{Aperture Photometry}\label{aperturephotometry}

We started by performing simple (stationary) aperture photometry on the K2 target pixel files. The K2 target pixel files contain a time series of images of the immediate vicinity of each target on the focal plane. The typical size of these subimages, which are colloquially referred to as ``postage stamps,'' ranged from about 25~$\times$~25 pixels in Campaign 0 to 10~$\times$~10 pixels in Campaign 3. We used the astrometric information from the K2 FITS headers to identify the target in question and refined our estimate of its position by finding the brightest pixel within a 3 pixel radius of the astrometric prediction. We defined 20 different aperture masks of various sizes centered about this point, 10 of which were circles, and 10 of which were shaped like the known \Kepler\ pixel response function (PRF) at the position of each target on the focal plane. The images of very bright stars with \Kepler\ band magnitudes brighter than ${\rm Kp}=9.5$ often have large bleed trails, so the \Kepler\ PRF does not describe the shape of the stellar image very well. For these very bright stars, instead of shaping 10 apertures like the \Kepler\ PRF, we define apertures by finding contiguous regions of pixels registering levels of flux greater than various threshold values (spaced between 0.2\% and 10\% of saturation values). 

After defining the photometric apertures, we calculated the median flux value of each image outside of a 5 pixel radius from the position of the target. In Campaign 0-2 data releases, the \Kepler\ pipeline did not perform any background subtraction of the pixel level data, so we subtracted this median value from each image. For the Campaign 3 data release, the \Kepler\ pipeline did perform background subtraction on the pixel level data, so we skipped this step. We then summed the flux contained within each of the 20 photometric apertures at each time stamp. In addition to calculating the flux within each aperture, we measured the centroid position of the target star at each timestamp. Following VJ14, we measured the centroid position in two ways: calculating the ``center of flux'', and fitting a Gaussian to the core of the PSF. 

\subsubsection{K2 Systematics Correction}\label{correction}

Raw K2 aperture photometry is dominated by systematic trends related to the motion of the spacecraft as its pointing slowly drifts due to solar radiation pressure and is corrected by thruster fires. Therefore, we perform processing that corrects for the systematic noise caused by \Kepler's motion. We follow a procedure similar to that described by VJ14, but modified for use on the datasets longer than the 9 day K2 engineering test. We begin by excluding datapoints taken during thruster firing events and calculating the distance traveled by \Kepler\ in the roll direction (hereafter referred to as ``arclength'') from measured image centroids. Instead of using each individual star's image centroids to calculate arclength like VJ14, we use the image centroids for one particular well behaved bright star for each campaign, which significantly improves data quality for faint ($Kp \gtrsim 15$) stars.

After measuring the motion of the spacecraft, we iteratively measure and remove the dependence of measured flux on image centroid position for light curves from each of the 20 apertures from Section \ref{aperturephotometry}, largely following the procedure of VJ14. The biggest difference between data from the K2 engineering test and Campaigns 0-3 is that during a typical 80 day K2 campaign, K2's pointing drifts significantly transverse to the back and forth rolling of the spacecraft. This means that the approximation made by VJ14 that the motion of the spacecraft is confined to one dimension breaks down on timescales longer than about 10 days. We overcome this by breaking each K2 campaign into shorter ``divisions,'' and in each of these divisions, independently decorrelating the spacecraft's pointing with measured flux. We decorrelate each division while simultaneously fitting a basis spline with breakpoints every 1.5 days through a longer segment of the time series in order to preserve low frequency variability like starspot modulation. Like VJ14, we model the dependence of flux on centroid position with a piecewise linear function. We excluded outliers when calculating the flat field in order to preserve transit-like events. An example of both the raw and systematics corrected K2 light curve of a planet candidate host star are shown in Figure \ref{sff}. 

We note that the light curves produced for Campaigns 0-3 (as described up to this point) are identical to the ones available for download from MAST as part of the K2SFF high level science product\footnote{\url{https://archive.stsci.edu/prepds/k2sff/}} (HLSP). All processing that is described after this stage is in addition to the processing applied to those general-purpose light curves.  

After we produced corrected light curves for each of the 20 different apertures, we selected one default aperture for the transit search by calculating the photometric precision for each light curve and selecting the best one. Unlike for the general purpose light curves archived in the K2SFF HLSP (and the default light curves selected there), before calculating the photometric precision, we removed low frequency variability by robustly fitting a basis spline to the light curve with breakpoints every 0.75 days and dividing it away. For stars with high amplitude starspot modulation, this can somewhat change the default aperture chosen, and yield a light curve better suited for planet detection.

\section{Transit Search, Triage, and Vetting}\label{vetting}

\subsection{Transit Search}\label{tps}

\begin{figure*}[t!]
\epsscale{1}
  \begin{center}
      \leavevmode
\plotone{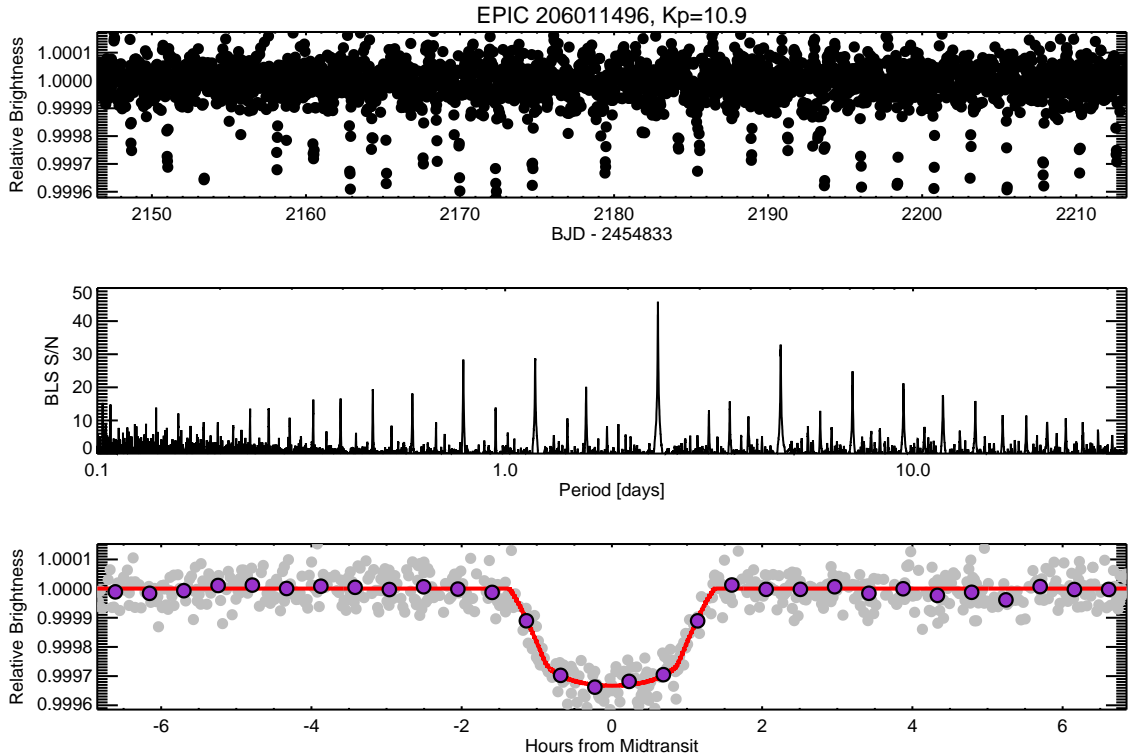}
\caption{Top: Flattened light curve of a 11th magnitude star with a transiting planet candidate. Middle: The BLS periodogram registers a strong peak at 2.37 days and harmonics of the 2.37 day peak. Bottom: Phase folded simultaneous-fit light curve (see Section \ref{transitfits}) and transit fit to the BLS detection.} \label{fig:flatbls}
\end{center}
\end{figure*}

After producing systematics--corrected light curves, we prepared them for our transit search. We began by removing low frequency variations from the light curve with the best photometric precision using a robustly fitted basis spline with breakpoints every 0.75 days, as done previously. We then removed outlier datapoints which, if ignored, can significantly decrease the sensitivity of the transit search. We removed 4--$\sigma$ upwards outliers, which are primarily caused by cosmic ray hits and asteroids passing in and out of the photometric apertures. In addition to these upwards outliers, our pipeline sometimes left in remnants of the pointing systematics which manifest as sparse downward outliers in the light curves. To limit the effect of these while being careful to avoid discarding astrophysical downward excursions (like transits), we performed limited downward-outlier exclusion. We did this by removing the two largest single-point downward excursions from each light curve. Transit events typically have durations longer than one long cadence datapoint, unless the planet's orbital period is very short, in which case there would be many more than two transits in the 35-80 days of K2 observations. Hence, only removing two single-point outliers should not significantly harm our ability to detect transits, and removing up to two spurious outliers improved our ability to detect transits. 

After flattening the light curve and performing outlier exclusion, we calculated a Box-Least-Squares (BLS) periodogram \citep{kovacs}, which is the heart of our transit search algorithm. We evaluated the periodogram over periods ranging from 2.4 hours to one half the total length of observations (which differs between the various K2 campaigns). We tested periods spaced such that: 

\begin{equation}
\frac{\Delta P}{P} = \frac{D}{N \times T_{\rm tot}}
\end{equation}

\noindent where $\Delta P$ is the spacing between successive periods evaluated in the periodogram, $P$ is each period tested, $D$ is the expected duration of a transit at each given period, $T_{\rm tot}$ is the total duration of K2 observations, and $N$ is the factor by which we oversampled the periodogram. Using a simplified expression for transit duration ($R_p << R_\star << a$ and transit impact parameter $b = 0$) and substituting in Kepler's third law, we can rewrite this as: 

\begin{equation}
\Delta P = \left[ \frac{P \times (365.25 {\rm~days})^{2}}{\pi^3 \times \rho_\star / \rho_\odot \times 215^{3}} \right]^{1/3} \frac{P}{N \times T_{\rm tot}}
\end{equation}

\noindent where $\rho_\star$ is the mean stellar density of the planet host, $\rho_\odot$ is the mean density of the sun, and 215 is the Earth's semimajor axis divided by the radius of the sun. This is equivalent to:

\begin{equation}
\Delta P = \frac{\rm 82~seconds}{N}\left(\frac{P}{\rm 1~day} \right)^{4/3} \left(\frac{\rho_\star}{\rho_\odot} \right)^{-1/3} \left(\frac{T_{\rm tot}}{\rm 80~days} \right)^{-1} 
\end{equation}

\noindent For our transit search, we evaluated $\Delta P$ and calculated periods to test using $\rho_\star$~=~6.5 $\rho_\odot$ and N~=~2 to ensure good sampling for planets orbiting stars much smaller and denser than the Sun. We also ensured that the duty cycle of the transits was at least three times larger than the size of each bin in phase tested by the BLS. 

After calculating the BLS periodogram, we took the output and normalized it in several ways. First, we noticed that at long periods, the baseline level of the BLS power ($P_{\rm BLS}$) typically rose because spurious detections were more likely. To prevent many spurious detections at long periods, we isolated and subtracted away the changing baseline level. We divided the BLS power spectrum into bins in period, calculated the median power measurement in each bin, interpolated between those median power measurements, and subtracted the interpolated baseline away. Then, we calculated the median absolute deviation (MAD) of the BLS power measurements, and normalized them to obtain the signal-to-noise ($S/N$) ratio of each BLS peak by calculating: 

\begin{equation}
S/N = P_{\rm BLS}/ ({\rm~MAD}/0.67)
\end{equation}

\noindent where 0.67 is the ratio between the MAD and standard deviation of a random normal distribution. Through trial and error, we determined that a signal-to-noise threshold of $S/N = 9$ struck a good balance between sensitivity to shallow transits and a manageable number of spurious detections.

Upon calculating a BLS periodogram which has at least one peak with $S/N > 9$ and a positive depth (that is, the star does not brighten), we performed some automatic tests. First, we rejected any detections with a duration greater than 20\% of the detected period and any detections with only one datapoint within the nominal transit window. In these cases, we automatically attempted to remove the data near the detected transits and recalculated the BLS periodogram. If a detection passed these tests, we then checked to see if one outlier datapoint was strongly influencing the depth of the putative transit. If removing the lowest datapoint changed the measured transit by more than 50\%, we removed that datapoint and recalculated the BLS periodogram, at which point the next detection began the process again. 

If, on the other hand, the detection passed these tests, then we promoted it to the status of ``Threshold Crossing Event'' (TCE). We fitted a \citet{mandelagol} transit model to the TCE light curve, and recorded the best-fit parameters. Then, we removed the datapoints within 1.5 transit durations of midtransit, and searched the light curve again with BLS. This process allowed us to detect systems of multiple transiting planets. 

We performed this type of transit search on all light curves corresponding to Guest Observer targets for each campaign. Calculating the BLS periodogram is the most computationally intensive part of the process, and each periodogram of a full length 80 day campaign took between 30 seconds and one minute per processor core. We show an example flattened light curve, BLS periodogram, and fitted transit in Figure \ref{fig:flatbls}.

\subsection{Triage and Vetting}

Our automatic transit search on all K2 data identified several thousand threshold crossing events. The majority of these TCEs, however, are not the signature of transiting exoplanets. In addition to genuine transiting planet candidates, TCEs can be caused by eclipsing binary stars, background eclipsing binary stars, stellar pulsations or other activity, and instrumental artifacts/uncorrected systematics left in the light curves. In this subsection, we describe the steps we have taken to filter out false positives and identify planetary candidates.  

\subsubsection{Triage}

After our pipeline produced a list of TCEs for each K2 campaign, we began the process of sorting genuine planet candidates from other events by performing triage. In this step, we quickly assessed whether the TCEs were spurious detections, data artifacts, or possible planet candidates by visual inspection of each candidate's transit light curve (both as a full time series and folded on the TCE period and ephemeris. During triage, we removed almost all signals caused by stellar oscillations, pulsations, or activity, many eclipsing binaries with significantly different primary and secondary eclipse depths, and most of the spurious detections caused by systematic effects or artifacts left in the data from the K2 roll dependent systematics. We also removed all BLS detections of a single event (such as single eclipses or transits of a long period system). During this stage, we were somewhat liberal in choosing which TCEs were allowed to pass to the more stringent vetting described in Section \ref{strictvetting}. 

\subsubsection{Light Curve and Pixel Vetting}\label{strictvetting}

After triage, we more closely examined the K2 light curve and pixel level data of each planet candidate to identify less obvious false positives. The tests we performed generally fell into two categories: tests designed to identify astrophysical false positives, and tests designed to identify instrumental false positives. At this stage, we produced postscript format diagnostic plots for each of our planet candidates, which are available on the ExoFOP-K2 website\footnote{\url{https://cfop.ipac.caltech.edu/k2/}}. 

\subsubsection{Identifying Instrumental False Positives}

We tested each of our surviving TCEs to identify spurious signals caused by imperfect systematics correction and other instrumental effects. While instrumental effects could cause spurious transits in \Kepler\ data \citep{koi6}, the problem is greater in K2 because of the spacecraft's unstable pointing. We performed three main tests. First, we checked to see if the points found to be ``in transit'' by BLS search tended to occur while \Kepler\ was pointed at a particular position along its roll, that is, if these points happened at particular values of arclength. In particular, the ``in transit'' points of many false positives were clustered at the far ranges of \Kepler's roll in arclength, where the K2 flat field is not well constrained. We also checked to see if the transit-like events were detectable in the light curves produced from other apertures besides the default one we chose in Section \ref{correction}. Spurious transit signals caused by improper removal of K2 roll systematics often differ significantly between the various apertures we considered. For bright stars ($Kp \lesssim 9.5$) with large bleed trails, we ignored any differences between photometric apertures, as any differences are likely due to the saturation of the CCD. Finally, we inspected each individual transit. The individual events of a spurious detection caused by instrumental systematics do not necessarily have consistent transit depths or shapes.

\subsubsection{Identifying Astrophysical False Positives}

Some transit-like events can be caused by astrophysical phenomena other than transiting exoplanets, including the case where the target star is an eclipsing binary star, a background source in the photometric aperture is an eclipsing binary, or a background source in the photometric aperture is a transiting planet host. These false positive sources are relatively well understood \citep{blender, morton2, morton12}, and were a concern in the original \Kepler\ mission as well \citep{batalha}. 

We followed \citet{batalha} and identified astrophysical false positives by searching for motion of flux centroids, high amplitude beaming or ellipsoidal variations in the light curve, and detections of secondary eclipses. Unlike the original \Kepler\ mission, K2's pointing is not stable on short timescales, so we performed our analysis of image centroids following \citet{foremanmackey}, and calculated the distance in centroid movement transverse to the roll of the spacecraft. This analysis is also affected by the motion of \Kepler\ transverse to its roll in long campaigns, so we calculated the distance transverse to the roll of the spacecraft in the same short ``divisions'' we used in Section \ref{correction}. We identified these astrophysical false positives with visual inspection of the image centroid positions and phase-folded light curves.

We promoted TCEs that survived both types of vetting to ``planet candidate'' status. We present our list of planet candidates in Table \ref{tab:planetparams}. Following the most recent editions of the \Kepler\ Objects of Interest (KOI) catalog, we don't necessarily consider candidates with very deep transits to be false positives, even though candidates with transit depths larger than $\simeq$ 5\% are very likely some sort of eclipsing binary. However, if candidates with transit depths greater than 5\%  survive our vetting, we list them separately from the rest of our planet candidates, at the end of Table \ref{tab:planetparams}. 

\section{Transit Light Curve Analysis}\label{transitfits}

After identifying planet candidates in K2 light curves with systematics corrected using our general purpose pipeline, we re-derived the K2 systematics correction while simultaneously fitting for the transit properties and shapes. We adopted a similar procedure to that described in \citet{hip116454} and \citet{becker}, and modeled the flat field as splines in arclength (with breakpoints spaced roughly every 0.25 arcseconds in arclength), the transits of all planet candidates orbiting each star with \citet{mandelagol} models, and low frequency variability in the light curve as a spline in time (with breakpoints spaced every 0.75 days). Like in our general purpose pipeline, we broke up the light curve into several divisions, and allowed the flat field within each division to vary separately. We performed the fit using a Levenberg-Marquardt least squares minimization algorithm \citep{mpfit}. Before fitting the light curves, we identified by eye cases where our pipeline detected the wrong orbital period for a particular candidate, and corrected the error. Period misidentifications typically happened for planet candidates with orbital periods longer than the longest period we searched (which was equal to half the length of observations).

The output of these fits is both transit parameters and a  flat field model. We then used the best-fit flat field model to produce a new light curve, and re-fit the transit parameters and low frequency variability using Levenberg-Marquardt while holding the flat field fixed. We report these parameters in Table \ref{tab:planetparams}. We also make the simultaneously fitted light curves available for download on the ExoFOP-K2 website\footnote{\url{https://cfop.ipac.caltech.edu/k2/}}. 

\section{Host Star Parameters}\label{stellarparams}

\subsection{Photometric Classification}\label{phottemp}

Estimating stellar parameters is difficult for K2 host stars because unlike the original \Kepler\ mission, which had an input catalog produced using uniform and high quality photometry \citep{kic}, K2's Ecliptic Plane Input Catalog (EPIC\footnote{\url{ https://archive.stsci.edu/k2/epic.pdf}}) was compiled from heterogenous sources of archival photometry. Because of this, the quantity and quality of available photometry for any given star is highly variable. Despite these challenges, attempts are underway to estimate stellar parameters from archival photometry (D. Huber, private communication 2015). In this work, we take a greatly simplified approach to give rough estimates of stellar parameters from photometry. Our approach is to assume that all of the stars we consider are main sequence dwarfs and estimate their parameters under that assumption. Then, using the stars' reduced proper motion, we flag some stars as ``Possible Giants.'' 

We start by querying the EPIC catalog for each of our candidate host stars. The EPIC draws photometry from the Hipparcos catalog \citep{hipparcos}, the Tycho-2 catalog \citep{tycho}, the Fourth US Naval Observatory CCD Astrograph Catalog (UCAC-4) survey \citep{ucac}, the Two Micron All Sky Survey (2MASS) \citep{twomass}, and the Sloan Digital Sky Survey \citep[SDSS, ][]{sdss}. The EPIC also contains proper motion and parallax information from these sources where available. The EPIC does not include photometry or proper motions for our Campaign 0 targets, so for these stars, we queried the 2MASS and UCAC4 catalogs for photometry and proper motions using Vizier. 

We then estimated the stellar effective temperature for each target from the star's colors using either empirical relations from \citet{boyajianextended}, \citet{casagrande}, or \citet{irfm}, or by interpolating color temperature relations from Eric Mamajek's spectral type table\footnote{\url{http://www.pas.rochester.edu/~emamajek/EEM_dwarf_UBVIJHK_colors_Teff.txt}} \citep{mamajek}. The method used to estimate stellar temperature depended on the available photometry. By default, we used the \citet{boyajianextended} $V-K$ relation. When one or both of the photometric measurements required were unavailable, we attempted to use (sequentially) the, \citet{boyajianextended} $B-V$ relation, the \citet{boyajianextended} $g-r$ relation, and when only infrared photometric measurements were unavailable, we used the \citet{irfm} $J-K$ relation. When the colors fell outside of any of the calibrated ranges of these relations, we instead used the Mamajek spectral type tables to estimate temperature, except for very red stars with both $V$ and $K$ magnitudes, when we used the \citet{casagrande} $V-K$ relation. When the $g-r$ colors fell outside the \citet{boyajianextended} calibrated range, we converted the $g$ and $r$ SDSS colors to Johnson V and Cousins R using the transformations from \citet{jordi}, and for stars with only infrared colors, we interpolated the Mamajek $H-K$ colors instead of $J-K$, because in M-dwarfs, $J-K$ is more sensitive to metallicity \citep{newton}. Like \citet{boyajianextended}, for their $V-K$ relation, we converted from 2MASS $K$ magnitudes to Bessel and Brett K magnitudes using the updated transformation\footnote{\url{http://www.astro.caltech.edu/~jmc/2mass/v3/transformations/}} from \citet{carpenter}, and converted from the Bessel and Brett system to the Johnson system using the transformation in \citet{besselbrett}. We list both the estimated effective temperatures for each planet candidate and the method used to produce the estimates in Table \ref{tab:stellar}. We make no attempt to correct for interstellar reddening, even though it affects some stars, in particular stars in Field 2. 

Once we had an estimate of effective temperature, we used the \citet{boyajiankm} temperature/radius relationship to estimate stellar radii. This relationship is calibrated between stellar effective temperatures of 3200 K and 5778 K. For stars with temperatures between 5778 K and 7000 K, we used the temperature/radius relation calculated by \citet{cc}, and for stars with temperatures greater than 7000 K or less than 3200 K, we used interpolated values from the Mamajek spectral type table to estimate stellar radii. Our radii are also listed in Table \ref{tab:stellar}. 

We also tabulated the measured proper motions for stars where available. We calculated the reduced proper motion $H_J$ in J-band using:

\begin{equation}\label{reducedpropermotion}
H_{J} = 5 + J + 5\log_{10}{PM_{\rm tot}}
\end{equation}

\noindent where $J$ is the 2MASS J-band apparent magnitude and $PM_{\rm tot}$ is the star's total proper motion in arcseconds~yr$^{-1}$. We used the reduced proper motion cut defined by \citet{cc} to identify possible giants amount the planet candidate host stars, and note those in Table \ref{tab:stellar}. 

\subsection{High Resolution Spectroscopy}\label{spectroscopy}

We observed many of the brighter (typically $Kp < 13$) candidate planet host stars with the high resolution Tillinghast Reflector Echelle Spectrograph (TRES) on the 1.5 meter telescope at Fred L. Whipple Observatory (FLWO) on Mt. Hopkins, Arizona. TRES is a fiber-fed optical echelle spectrograph which conducts observations with a spectral resolving power of $R = \lambda/\Delta\lambda = 44$,$000$. TRES can achieve radial velocity precision of approximately 10 \ms\ for high signal-to-noise observations, and the spectra can be used for precise stellar characterization.  

We observed \nst\ stars with TRES at least once, collecting a total of \nspec\ spectra, and extracted the spectra using the procedure described in \citet{buchhave10}. For typical twelfth magnitude targets, a 30 minute exposure yielded a signal-to-noise ratio of about 30 per resolution element. It is possible to measure precise stellar parameters for spectra of this signal-to-noise ratio using the Stellar Parameter Classification \citep[SPC,][]{buchhave, buchhave14} method. SPC works by cross correlating against a suite of synthetic stellar spectra based on Kurucz atmosphere models \citep{kurucz}, varying the stellar effective temperature $T_{\rm eff}$, metallicity [M/H], surface gravity $\log{g}$, and projected rotational velocity $v\sin{i}$.  We analyzed our TRES spectra with SPC where possible, and report the resulting spectroscopic parameters in Table \ref{tab:tres}. We only report SPC results for stars with $T_{\rm eff} > 4200$, $\log{g} > 4.2$ and for observations with a signal to noise ratio per resolution element greater than 25 and a normalized cross correlation peak height greater than 0.9. 

Of the \nst\ targets with at least one TRES observation, \nmulttres\ targets have more than one observation. For these stars, we measured precise relative radial velocities by cross correlating each spectrum (using multiple echelle orders) with the highest signal-to-noise observation. In these cases, we were able to compare the radial velocities of the spectra to detect or rule out massive companions. We report the scatter in radial velocity in Table \ref{tab:tres}. The individual relative radial velocity measurements are reported in \ref{tab:vels} and are also available on the ExoFOP-K2 website\footnote{\url{https://cfop.ipac.caltech.edu/k2/}}. 

In Table \ref{tab:vels} we report our velocity results for each individual TRES observation in a format suitable for combining with future follow-up velocity observations.  When there are rich data sets of relative radial velocities, it is sometimes feasible to combine data sets by allowing the offset between velocity zero points to be a free parameter.  This can be dangerous when there is no overlap in the time coverage between the data sets.  Therefore, we report absolute velocities on a system as close as possible to that defined by the IAU Radial Velocity Standard Stars \citep[e.g.][]{rvstandards}, based on extensive observations of several primary standards spread around the sky, and checked with observations of the solar spectrum during the day and minor planets during the night.  For our absolute velocities we use just the TRES order containing the Mg b triplet near 519 nm and run grids of correlations against model spectra drawn from the Kurucz/CfA library of model spectra.  We report the velocity from the synthetic template that gave the highest peak correlation value, after correcting by $-0.61$ \kms. This correction is mostly due to the fact that the CfA library does not include the gravitational redshift of the star.  For the Sun, this correction would be $-0.64$ \kms, close to the measured correction. We also report a rough estimate of $v\sin{i}$ from the equatorial rotational velocity of the library spectrum that produced the highest peak correlation value. These estimated rotational velocities can be useful for selecting the best (most slowly rotating) host stars for precise radial velocity observations. Table \ref{tab:vels} also reports the individual relative velocities from the multi-order velocity analysis and estimates of their internal error based on the scatter of the velocities from the different echelle orders.

\section{Discussion}\label{discussion}

\subsection{Characteristics of Planet Candidate Sample}

\begin{figure}[t!]
\epsscale{1}
  \begin{center}
      \leavevmode
\plotone{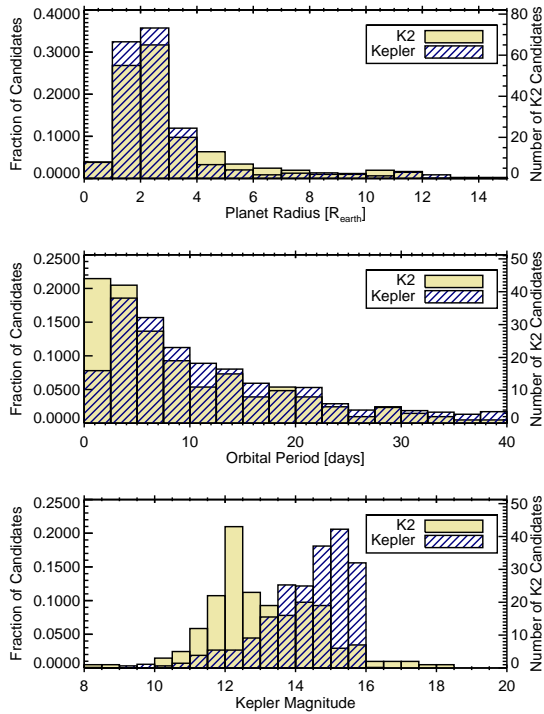}
\caption{Histograms of our K2 planet candidate sample (solid yellow fill) compared with planet candidates from the first four months of \Kepler\ observations (blue diagonal lines). Top: Histogram of planetary radii in our candidate sample and the KOI sample. Middle: Histogram of the orbital periods in our planet candidate sample and the KOI sample Bottom: Histogram of \Kepler-band magnitudes of our planet candidate host stars and the KOI sample. In general, the characteristics of the \Kepler\ and K2 planet samples are quite similar. The K2 host stars are typically brighter than the \Kepler\ host stars, and there are significantly fewer short-period candidates in the \Kepler\ sample, possibly due to pipeline completeness effects.} \label{histograms}
\end{center}
\end{figure}

A major benefit of K2 is the fact that because it looks at many different fields, it can observe more bright stars than were possible in the original \Kepler\ mission. Bright stars hosting small transiting planets are important for learning about the masses and compositions of small exoplanets through radial velocity measurements \citep{marcy, dressing}, and for learning about planetary atmospheres through transit spectroscopy \citep{knutson, kreidberg}. 

We have compiled histograms of planet radius, orbital period, and \Kepler-band magnitude of both our K2 planet candidates and planet candidates from the first four months of the \Kepler\ mission \citep{koi2} in Figure \ref{histograms}. To compare the two datasets, we restricted the histograms to plot those candidates with transit depths less than 5\% and orbital periods less than 40 days. We find that like the candidates from the \Kepler\ mission, the majority of our K2 candidates are small, with about 50\% having planetary radii between 1.6 and 4 \rearth. The \Kepler\ catalog has a slightly larger proportion of small candidates, likely due to the slightly longer 4-month observing baseline and \Kepler's superior photometric precision. The distribution of planet candidate orbital periods found by \Kepler\ and K2 are typically similar as well except that \Kepler\ finds significantly fewer candidates at very short orbital periods (P \textless 2.5 days). We believe this difference is likely due to the \Kepler\ pipeline's insensitivity to planets with very short orbital periods. Before searching for transits, the \Kepler\ pipeline pre-whitens light curves using a sinusoidal harmonic filter \citep{tenenbaum}, which substantially decreases the \Kepler\ pipeline's ability to detect planets with orbital periods less than about 2 days \citep{christiansen1, christiansen2}. When searching for K2 planet candidates, we do not remove sinusoids and their harmonics, so our pipeline remains sensitive to the shortest period planets. Finally, we find that our K2 planet candidates, though fewer in number, are typically brighter than \Kepler\ planet candidates.

There are 26 planet candidates orbiting stars with \Kepler-band magnitudes brighter than 12, with sizes between 1 and 4 Earth radii. These candidates are well suited for precise radial velocity follow-up. There are 10 candidates with radii between 1.6 and 4 Earth radii (that is, likely to have gaseous envelopes) with K-band magnitudes brighter than 10, and transit depths greater than 0.1\%. These candidates could be suitable for atmospheric characterization. Finally, using our (admittedly rough) estimates of stellar and transit parameters, we find that there are 8 sub-Earth sized candidates, the smallest of which are about 0.75 times the size of the Earth.

\subsection{Comparison with Existing Catalogs and Detections}

\begin{deluxetable}{cccc}[t!]
\tablewidth{250pt}
\tablecaption{Median 6 Hour Photometric Precision \label{precisioncomp}}
\tablehead{
\colhead{$K_{p}$} & \colhead{K2 C1} &\colhead{K2 C3} &  \colhead{\Kepler}}
\startdata

10-11            & 27    & 18    & 18         \\
11-12            & 29    & 22    & 22         \\
12-13            & 39    & 34    & 30         \\
13-14            & 62    & 56    & 47         \\
14-15            & 125   & 115   & 81         \\
15-16            & 266   & 236   & 147        \\
\enddata
\tablecomments{These measurements are the median 6 hour photometric precision (as defined by VJ14) in parts per million for dwarf stars in Campaign 1 of K2, Campaign 3 of K2, and the original \Kepler\ mission (calculated using the PDCSAP\_FLUX light curves produced by the \Kepler\ pipeline.)}
\end{deluxetable}

Previous studies have already identified some of the candidates identified in our search. The largest transit search in K2 data to date was undertaken by \citet{foremanmackey}, who searched for candidates in all of the data taken during Campaign 1 and reported 36 candidate transit signals \citep[19 of which were statistically validated by][]{montet}. Out of the 36 candidate transit signals reported by \citet{foremanmackey}, we identify and include 31. Of the 5 we do not report, two (201929294 and 201555883) are spurious transit signals \citep[as identified by][]{montet}, and were not detected by our pipeline. One (201702477) was thrown out during triage because the light curve did not appear transit-like, one (201393098) was missed by our pipeline (likely due to a poor systematics correction, though the transits are evident in our phase-folded light curve), and one (201338508.02) was rejected because it showed an 8.5--$\sigma$ centroid offset during transit. EPIC 201338508 is an interesting case because it hosts another transiting planet (which showed no significant centroid offset during transit), and false positives in multi-candidate systems are thought to be rare \citep{lissauer}. However, given the difficulties in performing centroid analyses on K2 data, it is possible the centroid offset detected for 201338508.02 is spurious, and the candidate is in fact a genuine planet.

In total, our search returned 78 planet candidates from Campaign 1, including 48 with periods longer than 4 days, the shortest period searched by \citet{foremanmackey}. This means we detect 19 new candidates in the period range searched by \citet{foremanmackey}. The 25$^{\rm th}$ percentile transit depth of the 19 additional planets we detect is about 35\% smaller than the 25$^{\rm th}$ percentile transit depth of the 29 planets we detect in common with \citet{foremanmackey}, indicating that our pipeline is sensitive to shallow transits. 

We detect and list previously known planets discovered by ground based surveys and observed by K2: HAT-P-54 \citep[EPIC 202126849,][]{hatp54}, HAT-P-56 \citep[EPIC 202126852,][]{huang}, WASP-85 \citep[EPIC 201862715,][]{wasp85}, WASP-47 \citep[and its two additional transiting planets; EPIC 206103150,][]{wasp47, becker}, and WASP-75 \citep[EPIC 206154641,][]{wasp75}. We detect two previously announced systems of small planets around bright M-dwarfs: EPIC 201367065 \citep{crossfield} and EPIC 206011691 \citep{petigura}, and the previously announced hot Jupiter transiting EPIC 204129699 \citep{Grziwa}. We also detect the previously announced disintegrating objects transiting EPIC 201637175 \citep{roberto} and WD 1145+017 \citep{wd1145}. 

We make special note of the EPIC 201505350 (also designated K2-19) system. \citet{armstrong2} and \citet{foremanmackey} previously identified two transiting giant planet candidates around EPIC 201505350, close to a 3:2 mean motion resonance, and \citet{armstrong2} confirmed their planetary nature by detecting transit timing variations. In addition to recovering these two transit signals, we detect a roughly 1.25 \rearth\ planet candidate interior to these two planets in a 2.5 day orbital period. If the third candidate is in fact a genuine planet, it could be important in future dynamical studies of this system \citep[e.g.][]{barros, narita}. 

\subsection{Improvement to Campaign 3 Photometry}\label{c3precision}

\begin{figure*}[ht!]
\epsscale{1}
  \begin{center}
      \leavevmode
\plotone{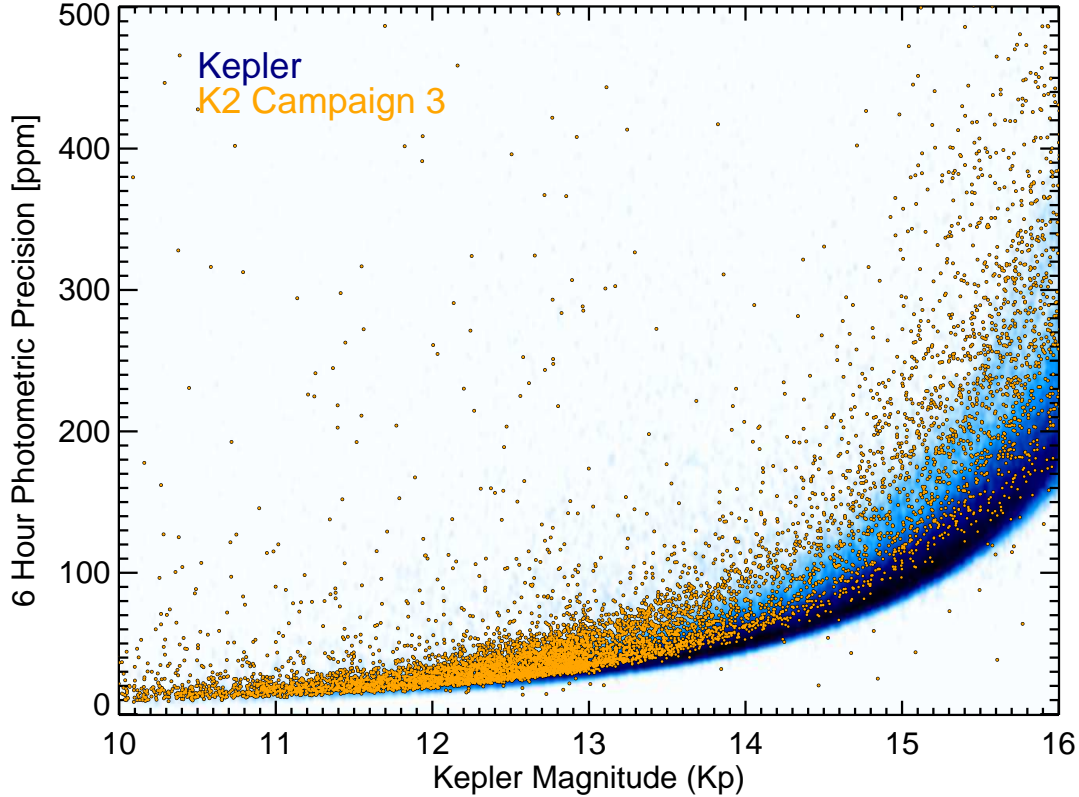}
\caption{Photometric precision of \Kepler\ stars (blue) compared with stars observed by K2 during Campaign 3 (orange circles). K2 attains almost equivalent photometric precision to \Kepler\ for stars brighter than $Kp \simeq 12.5$. There are a similar number of upwards outliers (high photometric variability stars) in \Kepler\ as in K2, indicating that the majority of those are showing astrophysical variability and not instrumental effects.} \label{k2vsk1}
\end{center}
\end{figure*}

Although the main point of this paper is to provide a list of planet candidates for future follow-up, we make note of a major improvement to the quality of K2 photometry in Campaign 3. Previous to Campaign 3, K2 photometry of bright stars exhibited a noise floor at the level of roughly 25 parts per million (ppm) per six hours \citep[see, for example, VJ14; ][]{aigrain, huangk2c1}. This noise floor is significantly worse than the noise floor in the original \Kepler\ mission of roughly 10 ppm per six hours. Early on in the K2 mission, it was believed that this noise floor was caused by the lower frequency of pointing adjustments in K2 compared to \Kepler. During the original \Kepler\ mission, the spacecraft adjusted its pointing every 10 seconds, but that frequency had been decreased to every 50 seconds in early K2 operations. After the pointing performance of the spacecraft in its K2 operating mode was deemed suitable, the \Kepler/K2 team increased the frequency of pointing adjustments from every 50 seconds to every 20 seconds starting in Campaign 3. 

The photometric precision of our Campaign 3 light curves appears to show that this change has led to a significant improvement to K2 photometric precision. In Table \ref{precisioncomp}, we summarize the median photometric precision for dwarf stars achieved by K2 during both Campaign 1 (before the increase in correction bandwidth), Campaign 3 (after the increase in bandwidth), and the median photometric precision achieved by \Kepler\ in its original mission. We also compare the photometric precision of individual stars in both \Kepler\ and K2 versus \Kepler\ magnitude in Figure \ref{k2vsk1}. For stars brighter than about $Kp \simeq 12.5$, K2's photometric precision is essentially equal to that of the original \Kepler\ mission (although there are still significantly more artifacts in corrected K2 data than in original \Kepler\ data). For stars fainter than $Kp \simeq 12.5$, the larger apertures required for K2 photometry still limit the typical precision to be somewhat worse than the original \Kepler\ mission.  

If the improvement in photometric precision in Campaign 3 still holds in future K2 campaigns, it will be easier to detect small planet candidates around the brightest stars observed by K2. For example, the improved photometric precision in Campaign 3 enabled the detection of shallow (275 ppm) transits in a relatively long (10 day) period around the 8th magnitude star EPIC 205904628 (HD 212657). K2 achieved a photometric precision of 12 ppm per six hours on this very bright star. This planet candidate would have been much more difficult to detect in earlier K2 campaigns. 

\section{Summary}

We have processed and searched data from the \Kepler\ space telescope in its K2 extended mission for transiting planet candidates. After searching through the first year of K2 data (Campaigns 0, 1, 2, and 3), we identify \nplanets\ planet candidates around \nsys\ stars. Many of these planet candidates orbit bright stars, suitable for atmospheric characterization and radial velocity follow-up. For many of the brighter planet host stars, we have obtained reconnaissance high resolution optical spectroscopy with the TRES spectrograph on the 1.5 meter telescope on Mt. Hopkins. We derive stellar spectroscopic parameters from the TRES data, and in some cases where we have multiple observations, we extract precise radial velocities to either detect or exclude stellar-mass companions. 

We have made all of our data products available online, including processed light curves, light curve and candidate vetting diagnostics, reduced spectra, radial velocities, and stellar parameters. We also note that a change to the K2 observing strategy starting with Campaign 3 has greatly improved K2's photometric precision for the brightest targets. This bodes well for future detections of small planets around bright stars, which are important for understanding the composition of small exoplanets. 

\acknowledgments
We thank Dave Charbonneau, Courtney Dressing, Xavier Dumusque, Dan Foreman-Mackey Mercedes Lopez-Morales, Ben Montet, Elisabeth Newton, and Roberto Sanchis-Ojeda for helpful discussions. We thank the referee, Drake Deming, for a thoughtful report with suggestions that improved the manuscript. We thank Chris Shallue for identifying a typographical error in Equation 3. We gratefully acknowledge the efforts of the entire \Kepler/K2 team who made this work possible. A.V. is supported by the NSF Graduate Research Fellowship, Grant No. DGE 1144152. DWL gratefully acknowledges support from the \Kepler\ mission under NASA cooperative agreement NNX13AB58A to the Smithsonian Astrophysical Observatory. We additionally acknowledge support from the John Templeton Foundation. The opinions expressed in this publication are those of the  authors and do not necessarily reflect the views of the John Templeton Foundation. This research has made use of NASA's Astrophysics Data System, the VizieR catalogue access tool, CDS, Strasbourg, France, and the NASA Exoplanet Archive (particularly ExoFOP-K2), operated by the California Institute of Technology, under contract with NASA under the Exoplanet Exploration Program. Some of the data presented in this paper were obtained from the Mikulski Archive for Space Telescopes (MAST). STScI is operated by the Association of Universities for Research in Astronomy, Inc., under NASA contract NAS5--26555. Support for MAST for non--HST data is provided by the NASA Office of Space Science via grant NNX13AC07G and by other grants and contracts.This paper includes data collected by the \Kepler/K2 mission. Funding for the \Kepler\ mission is provided by the NASA Science Mission directorate.

Facilities: \facility{Kepler, FLWO:1.5m}


\clearpage
\LongTables
\begin{deluxetable*}{ccccccccccccc}
\tablewidth{0pt}
\tabletypesize{\scriptsize}
\tablecaption{\label{tab:planetparams} Planet Candidate Parameters}
\tablehead{\colhead{ID} &
\colhead{RA} &
\colhead{Dec} &
\colhead{$K_p$} &
\colhead{Period} &
\colhead{$T_{0}$} &
\colhead{$T_{1-4}$} &
\colhead{$b$} &
\colhead{$\delta$} &
\colhead{$R_{p}/R_{\star}$} &
\colhead{$R_\star$} &
\colhead{$R_{\rm p, fit}$} &
\colhead{$R_{\rm p, \delta}$} 
\\
\colhead{} &
\colhead{} &
\colhead{} &
\colhead{} &
\colhead{{\em days}} &
\colhead{{\em BJD-2454833}} &
\colhead{{\em hours}} &
\colhead{} &
\colhead{\%} &
\colhead{} &
\colhead{$R_\odot$} &
\colhead{$R_\oplus$} &
\colhead{$R_\oplus$}}
\startdata
202059377.01 & 06:36:44.230 & +21:39:05.65 & 8.70 & 1.5771832 & 1939.583 & 3.96 & 1.00 & 2.49 & 0.2491 & 1.94 & 52.8 & 33.4 \\ 
202066192.01 & 06:20:59.388 & +26:07:31.37 & 14.80 & 3.2923818 & 1941.826 & 1.76 & 0.90 & 0.924 & 0.1053 & 1.87 & 21.5 & 19.6 \\ 
202066537.01 & 06:57:03.847 & +18:37:36.26 & 14.80 & 6.9055211 & 1942.097 & 3.65 & 0.89 & 2.65 & 0.1906 & 1.89 & 39.4 & 33.6 \\ 
202071289.01 & 06:20:43.942 & +24:54:15.62 & 11.00 & 3.0459383 & 1941.358 & 2.21 & 0.77 & 3.82 & 0.1997 & 0.89 & 19.4 & 19.0 \\ 
202071401.01 & 06:41:30.713 & +20:55:22.12 & 12.90 & 3.2385420 & 1939.226 & 1.27 & 0.39 & 0.0499 & 0.02037 & 0.69 & 1.52 & 1.67 \\ 
202072596.01 & 06:38:02.578 & +20:56:21.52 & 11.20 & 3.9792870 & 1939.894 & 4.82 & 0.88 & 3.50 & 0.2167 & 1.84 & 43.5 & 37.5 \\ 
202083828.01 & 06:16:49.579 & +24:35:47.08 & 14.00 & 14.567025 & 1942.164 & 4.66 & 0.00 & 0.255 & 0.04567 & 0.51 & 2.52 & 2.79 \\ 
202086968.01 & 06:28:01.351 & +16:23:19.86 & 12.40 & 1.8490876 & 1939.183 & 3.12 & 0.97 & 2.74 & 0.2448 & 0.96 & 25.5 & 17.3 \\
\enddata
\tablecomments{ In this table, $b$ is the transit impact parameter, $T_{1-4}$ is the duration of the transit between first and fourth contact, $T_0$ is the time of transit, $R_p$ is the planetary radius, and $\delta$ is the depth of the transits. We estimate the planetary radius in two ways and report both values in this table. The column labeled $R_{\rm p, fit}$ is calculated by multiplying the best-fit $R_p/R_\star$ values with the stellar radius. The column labeled $R_{\rm p, \delta}$ is calculated by multiplying $\sqrt{\delta}$ with the stellar radius. Our best-fit parameters can sometimes converge to unphysical high impact parameter values, in which case $R_{\rm p, \delta}$ is a more reliable estimator of the planetary radius.}
\end{deluxetable*}

\clearpage
\LongTables
\begin{deluxetable*}{ccccccccccc}
\tablewidth{0pt}
\tabletypesize{\scriptsize}
\tablecaption{\label{tab:stellar} Stellar Parameters for Planet Candidate Host Stars}
\tablehead{\colhead{EPIC} &
\colhead{$K_p$} &
\colhead{RA} &
\colhead{Dec} &
\colhead{Campaign} &
\colhead{$T_{\rm eff,~phot}$} &
\colhead{$R_{\rm\star,~phot}$} &
\colhead{$\rm PM_{\rm tot}$} &
\colhead{Reduced PM} &
\colhead{Possible} &
\colhead{$N_{\rm TRES}$} \\
\colhead{} &
\colhead{} &
\colhead{} &
\colhead{} &
\colhead{} &
\colhead{$K$} &
\colhead{$R_{\odot}$} &
\colhead{{\em mas~yr$^-1$}} &
\colhead{$H_{J}$} &
\colhead{Giant?} &
\colhead{}}
\startdata
202059377 & 8.70 & 06:36:44.230 & +21:39:05.65 & 0 & 8210$^a$ & 1.94 & 3.7 & -3.66 &  &  \\
202064253 & 12.30 & 06:06:01.015 & +22:53:30.34 & 0 & 6205$^a$ & 1.29 & 3.2 & -1.09 &  &  \\
202066192 & 14.80 & 06:20:59.388 & +26:07:31.37 & 0 & 7824$^a$ & 1.87 & 1.2 & -0.21 &  &  \\
202066537 & 14.80 & 06:57:03.847 & +18:37:36.26 & 0 & 6722$^d$ & 1.89 & 24.8 & 5.71 &  &  \\
202071289 & 11.00 & 06:20:43.942 & +24:54:15.62 & 0 & 5529$^a$ & 0.89 & 17.5 & 0.91 &  &  \\
202071401 & 12.90 & 06:41:30.713 & +20:55:22.12 & 0 & 4446$^a$ & 0.69 & 271.1 & 7.95 &  &  \\
202071635 & 10.20 & 06:14:17.234 & +18:37:37.27 & 0 & 7254$^a$ & 1.81 & 1.9 & -3.92 &  &  \\
202072596 & 11.20 & 06:38:02.578 & +20:56:21.52 & 0 & 7430$^a$ & 1.84 & 5.6 & -0.40 &  &  \\
\enddata
\tablecomments{$K_p$ is the \Kepler\ band magnitude of the host star. $T_{\rm eff,~phot}$ is our photometric estimate of the stellar effective temperature, obtained following the procedure outlined in Section \ref{phottemp}. $R_{\rm\star,~phot}$ is our estimate of the stellar radius based on the photometric effective temperature. $\rm PM_{\rm tot}$ is the total proper motion of the host star. $H_{J}$ is the reduced proper motion of the star, calculated using Equation \ref{reducedpropermotion}. The ``Possible Giant'' column is a binary flag noting possible giant host stars based on reduced proper motion. $N_{\rm TRES}$ is the number of spectroscopic observations we obtained with TRES. These observations are summarized further in Table \ref{tab:tres}.}
\tablenotetext{$a$}{$T_{\rm eff}$ derived using the \citet{boyajianextended} V-K relation.}
\tablenotetext{$b$}{$T_{\rm eff}$ derived using the \citet{boyajianextended} B-V relation.}
\tablenotetext{$c$}{$T_{\rm eff}$ derived using the \citet{boyajianextended} g-r relation.}
\tablenotetext{$d$}{$T_{\rm eff}$ derived using the \citet{irfm} J-K relation.}
\tablenotetext{$e$}{$T_{\rm eff}$ derived by interpolating the Mamajek H-K spectral type tables.}
\tablenotetext{$f$}{$T_{\rm eff}$ derived by interpolating the Mamajek V-R spectral type tables.}
\tablenotetext{$g$}{$T_{\rm eff}$ derived using the \citet{casagrande} V-K relation}

\end{deluxetable*}

\clearpage
\LongTables
\begin{deluxetable*}{cccccccccc}
\tablewidth{0pt}
\tabletypesize{\scriptsize}
\tablecaption{\label{tab:tres} Summary of Spectroscopic Observations with TRES}
\tablehead{\colhead{EPIC} &
\colhead{$N_{\rm TRES}$} &
\colhead{$T_{\rm eff, spec}$} &
\colhead{$\log{g}$} &
\colhead{[M/H]} &
\colhead{$v\sin{i}$} &
\colhead{Absolute RV} &
\colhead{stdev$_{\rm RV}$} &
\colhead{$\sigma_{\rm RV}$} &
\colhead{Notes}
\\
\colhead{} &
\colhead{} &
\colhead{$K$} &
\colhead{} &
\colhead{} &
\colhead{\kms} &
\colhead{\kms} &
\colhead{\ms} &
\colhead{\ms} &
\colhead{}}
\startdata
201295312 & 1 &  &  &  &  &  &  &  & g \\
201324549 & 2 &  &  &  &  &  &  &  & b,e \\
201367065 & 3 &  &  &  &  &  & 46 & 46 & d \\
201384232 & 1 & 5617$~\pm~$53 & 4.67$~\pm~$0.10 & -0.12$~\pm~$0.08 & 3.1$~\pm~$0.5 & 7.82$~\pm~$0.10 &  &  &  \\
201403446 & 2 &  &  &  &  &  & 25 & 47 & g \\
201441872 & 1 & 5450$~\pm~$50 & 4.54$~\pm~$0.10 & -0.13$~\pm~$0.08 & 3.5$~\pm~$0.5 & -1.51$~\pm~$0.10 &  &  &  \\
201460826 & 1 &  &  &  &  &  &  &  & g \\
201505350 & 2 & 5391$~\pm~$50 & 4.60$~\pm~$0.10 & -0.03$~\pm~$0.08 & 2.6$~\pm~$0.5 & 7.41$~\pm~$0.10 & 155 & 30 &  \\
\enddata
\tablecomments{ $N_{\rm TRES}$ is the number of spectroscopic observations. The output parameters from SPC are: $T_{\rm eff, spec}$, the stellar effective temperature; $\log{g}$, the stellar surface gravity; [M/H], the metallicity; and $v\sin{i}$, the projected rotational velocity. The observed scatter in radial velocities is listed in the column labeled stdev$_{\rm RV}$, while the typical measurement uncertainty is listed in the column labeled $\sigma_{\rm RV}$. An observed scatter in radial velocity larger than the measurement uncertainties could indicate real radial velocity variations.  }

\tablenotetext{$a$}{One exposure is significantly weaker than others, so interpret multi-order velocity results are not reported.}
\tablenotetext{$b$}{Composite spectrum, so SPC and multi-order velocity results are not reported.}
\tablenotetext{$c$}{Moonlight contamination, so interpret results with caution. SPC and multi-order velocities may not be reported depending on the contamination.}
\tablenotetext{$d$}{Cool star with $T_{\rm eff, spec} < 4200$, so SPC results are not reported.}
\tablenotetext{$e$}{Weak CCF with the normalized cross correlation peak height \textless 0.9, so SPC results are not reported.}
\tablenotetext{$f$}{Weak exposure with S/N per resolution element less than 25, so SPC results are not reported}
\tablenotetext{$g$}{Surface gravity $\log{g} < 4.2$, so SPC results are not reported.}

\end{deluxetable*}

\clearpage
\LongTables
\begin{deluxetable*}{cccccccc}
\tablewidth{0pt}
\tabletypesize{\scriptsize}
\tablecaption{\label{tab:vels} List of Individual Observations with TRES}
\tablehead{\colhead{EPIC} &
\colhead{BJD$_{\rm UTC}$} &
\colhead{Absolute RV} &
\colhead{Relative RV} &
\colhead{$v\sin{i}$} &
\colhead{SNR$_{e}$} &
\colhead{Exposure time} &
\colhead{Notes}
\\
\colhead{} &
\colhead{} &
\colhead{\kms} &
\colhead{\ms} &
\colhead{\kms} &
\colhead{} &
\colhead{s} &
\colhead{}}
\startdata
201295312 & 2457046.96711 & 44.54 &  & 4 & 48.7 & 1800 & \\
201324549 & 2457026.05794 & -13.17 &  & 12 & 22.9 & 600 &  composite spectrum- triple\\
201324549 & 2457118.89639 & -13.24 &  & 10 & 31.2 & 1200 &  composite spectrum- triple\\
201367065 & 2457027.03626 & 30.48 & -9.7 $~ \pm ~$ 37.0 & 4 & 23.4 & 1200 & \\
201367065 & 2457081.93418 & 30.31 & 73.7 $~ \pm ~$ 64.4 & 4 & 13.9 & 1350 & \\
201367065 & 2457092.94708 & 30.38 & 0.0 $~ \pm ~$ 37.0 & 2 & 25.2 & 1200 & \\
201384232 & 2457144.68312 & 7.82 &  & 2 & 27.4 & 1200 & \\
201403446 & 2457050.92825 & 18.91 & -34.8 $~ \pm ~$ 47.1 & 6 & 39.6 & 900 & \\
\enddata
\tablecomments{Absolute RV is measured from the cross correlation of the TRES order containing the Mg b triplet near 519 nm, as described in Section \ref{spectroscopy}. The accuracy of these measurements is perhaps 100 \ms, mostly limited by systematic errors in establishing the transfer to the IAU Radial Velocity Standard Stars. Relative RV is measured from the multi-order velocity analysis described in Section \ref{spectroscopy}, and can have a precision as good as about 15 \ms. The column labeled $v\sin{i}$ gives the equatorial rotational velocity of the model spectrum producing the highest cross-correlation peak, and is only a rough estimate of $v\sin{i}$. The values should not be trusted for rotational velocities slower than about 5 \kms. SNR$_e$ is the signal-to-noise ratio per resolution element in the continuum near 519 nm.}

\end{deluxetable*}

\clearpage

\end{document}